\documentclass[prb,twocolumn,groupedaddress,showpacs,showemail]{revtex4}
\usepackage{graphicx}

\begin{document}
\title{Transition rates for a $S \ge 1$ model
coupled to a phonon bath} 
\author{Kyungwha Park}\email{kyungwha@vt.edu}
\affiliation{
Department of Physics, Virginia Polytechnic Institute and
State University, Blacksburg, Virginia 24061 }
\date{\today}

\begin{abstract}
We investigate transition rates between different spin configurations
for $S \ge 1$ spins weakly coupled to a $d$-dimensional 
phonon bath. This study is motivated by 
understanding observed magnetization relaxation as a function of temperature 
in diverse magnetic systems such as arrays of magnetic nanoparticles and 
magnetic molecules. We assume that the magnetization of the spin system relaxes 
through consecutive emission or absorption of a single phonon. From a weak, linear
spin-phonon coupling Hamiltonian, we derive transition rates that would 
be used to examine dynamic properties of the system in kinetic Monte Carlo 
simulations. Although the derived phonon-assisted transition rates
satisfy detailed balance, in the case of two and three dimensional 
phonon baths, transitions between degenerate states are not allowed.
Thus, if there are no alternative paths along which the spin
system can relax, the relaxation time diverges. Otherwise, the system 
finds other paths, which leads to an increase in the
relaxation time and energy barrier. However, when higher-order 
phonon processes are included in the transition rates, it is found
that the system can reach the states which were inaccessible due to
the forbidden transitions. As a result, the system recovers some of 
the dynamic properties obtained using the Glauber transition rate.
\end{abstract}

\maketitle



\section{Introduction}

In many physical and chemical nanoscale systems ranging from semiconductor
quantum dots to arrays of magnetic nanoparticles or nanoscale magnetic 
molecules, dynamic properties play a crucial role in understanding the 
underlying physics and in designing systems of interest for practical 
applications. For example, the time evolution of quantum systems into 
decoherence needs to be fully understood in various local environments
in order to build scalable quantum computers. It is also important to 
investigate the spin-lattice relaxation time $T_1$ and the magnetization 
relaxation time for recently synthesized nanoscale magnetic systems 
to use them as information storage devices.

To study the dynamic properties of the systems discussed above, 
it is common to consider interactions of the systems with their 
environment. The environment is typically described as a heat bath, 
which has a much shorter relaxation time than the systems. 
Depending on the physical quantities to be calculated, one has 
to choose an appropriate bath. To understand decoherence mechanisms 
in quantum dots at low temperatures, one often considers an interaction 
between electron spins and a spin bath consisting of a large number 
of $S=1/2$ spins that mimic nuclear spins.\cite{ZHAN06,PROK00} To 
estimate the spin-lattice relaxation time in quantum dots, one should 
take into account electronic spins coupled to a phonon bath.\cite{MEUN07} 
In nanoscale magnetic systems such as arrays of magnetic nanoparticles, 
\cite{ZHAN01,KIM03} single-molecule magnets,\cite{POLI95,LEUE00,CHUD05} 
and single-chain magnets,\cite{CANE02,COUL04,KISH06} the effect of
nuclear spins is minimal, so a phonon bath becomes more relevant 
than a spin bath. For a single-molecule magnet embedded in a 
three-dimensional lattice, transition rates between different spin
configurations have been derived from coupling to the 
lattice via magneto-elastic coupling.\cite{POLI95,LEUE00,CHUD05,SOLO07} 
The magnetization relaxation time was estimated using 
these transition rates and quantum tunneling rates, being in 
good agreement with experimental data.\cite{LEUE00} 
For a nearest-neighbor interacting ferromagnetic Ising system, 
transition rates were derived from a weak, linear coupling of 
the system to a one-, two-, or three-dimensional phonon 
bath.\cite{parknov1,parknov2,parknov3} Using these phonon-assisted 
transition rates, 
kinetic Monte Carlo simulations were performed to measure the 
lifetime of the metastable state or magnetization relaxation 
time at low temperatures. The Monte Carlo simulations revealed 
that the dynamic properties obtained using the phonon-assisted 
transition rates greatly differ from those using other transition 
rates, such as Glauber \cite{GLAU63} or Metropolis.\cite{METR53} 
It is known that 
the Glauber transition rate can be derived from a coupling 
of a spin system to a fermionic bath.\cite{MART77}
Recently, it was shown that the soft Glauber transition rate 
\cite{RIKV02} requires different interpretation in the form of the
lifetime of the metastable state at low temperatures, 
although the energy barrier to reach equilibrium is the
same as that for the standard hard Glauber transition rate.\cite{PARK04}
Therefore, selection of a proper bath and relevant transition 
rates is critical in understanding dynamic properties. 
The transition rates derived by Park {\it et al.}
\cite{parknov1,parknov2,parknov3} (two- and three-dimensional
baths) and other groups \cite{LEUE00,CHUD05,SAIT00}
(three-dimensional bath) using coupling to a phonon bath, share a common feature 
that the rates become zero for degenerate states. Very recently, 
the derived phonon-assisted transition rates were used to 
examine the nanostructure of field-driven solid-on-solid 
interfaces.\cite{BUEN07} It was found that the phonon-assisted rates
provide significant differences from other types of transition rates, 
such as the Glauber dynamics.

In this study, targeting arrays of weakly interacting magnetic 
nanoparticles, single-molecule magnets, and single-chain magnets, 
we generalize the formalism used for an Ising system in 
Refs.\onlinecite{parknov1,parknov2,parknov3} to a $S \ge 1$ 
model on a lattice. Each spin in the model interacts via its
nearest neighbors and has easy-axis single-ion anisotropy.  
We assume that all spins in the model are weakly coupled to 
a phonon bath in $d=1,2$, or 3 dimensions. Considering that 
spin relaxation occurs through first-order one-phonon 
emission or absorption processes, we derive transition rates
for one-, two-, and three-dimensional phonon baths. In the 
cases of $d=2$ and 3, some transitions
or relaxation paths are inaccessible because transition rates 
between degenerate states vanish. This results in increasing
magnetization relaxation time and energy barrier to be overcome.
However, when higher-order phonon 
processes are included, other relaxation paths are opened up, 
leading to shortening of the relaxation time compared to
the first-order phonon processes. The formalism for the 
phonon-assisted transition rates is presented in Sec.II. 
The consequences of using the derived transition rates 
are discussed in the context of kinetic Monte Carlo simulations in 
Sec.III. Higher-order processes and their effects on dynamic
properties are presented in Sec.IV. The conclusion follows in
Sec.V.

\section{Formalism for phonon-assisted transition rates}

Although the current formalism can be applied to more general 
cases, we start with the following Hamiltonian for $N_s$ spins 
($S \ge 1$) on a lattice.
\begin{eqnarray}
{\cal H}_{\mathrm {sp}} &=& - J \sum_{\langle i,j \rangle} S_i^z S_j^z
-D \sum_{i=1}^{N_s} (S_i^z)^2 - H \sum_{i=1}^{N_s} S_i^z \:,
\label{eq:H_spin}
\end{eqnarray}  
where $S_i^z$ is the $z$ component of spin operator $\vec{S}$ at site $i$, 
$J(>0)$ is an exchange coupling constant between nearest-neighboring 
spins, and the first summation runs over all nearest neighbor pairs. 
A positive value of $J$ implies ferromagnetically coupled spins. 
$D(>0)$ is a single-ion magnetic anisotropy parameter determined by 
the spin-orbit coupling. A positive value of $D$ indicates that 
the magnetic easy axis of an individual spin is along the $\pm z$ 
axis. Notice that our convention on $D$ differs from other works.
$H$ is an external magnetic field applied to the spin system. 
We call the spin Hamiltonian Eq.~(\ref{eq:H_spin}) a $S \ge 1$ 
model. The eigenstates $|m \rangle$ of the spin Hamiltonian are 
\begin{eqnarray}
| m \rangle &=& |m_1 \rangle \bigotimes |m_2 \rangle \bigotimes \cdot
\cdot \cdot \cdot \bigotimes |m_{N_s} \rangle
\label{eq:eigenstates}
\end{eqnarray}
where $|m_i \rangle$ is the eigenstate of $S_i^z$ and $m_i=-S, -S+1, ..., S-1, S$. 
For $S=1$ and $D<0$, the spin Hamiltonian Eq.~(\ref{eq:H_spin}) 
is known as the Blume-Capel model,\cite{CAPE66,BLUM66} which was introduced to
describe features of the phase diagram of He$^3$-He$^4$ mixtures as well
as to understand a phase transition in UO$_2$.
The $S \ge 1$ model can be applied to the following magnetic
systems: arrays of weakly-interacting magnetic nanoparticles,\cite{ZHAN01,KIM03} 
nanoscale single-molecule magnets such as Mn$_{12}$ and Fe$_8$,\cite{POLI95,LEUE00,CHUD05} 
a Mn(III)$_2$Ni(II) single-chain magnet,\cite{COUL04,KISH06} and 
a Co ferrimagnetic compound.\cite{CANE02}

\begin{figure}
\includegraphics[angle=0,width=0.4\textwidth]{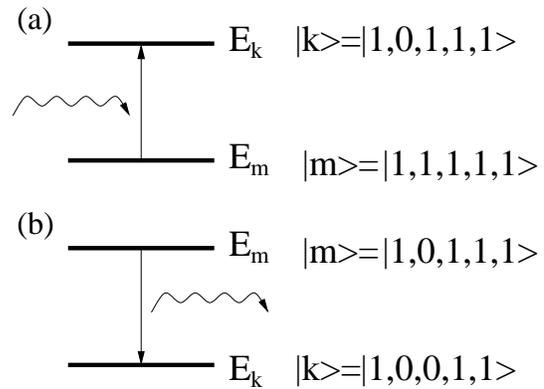}
\caption{Schematic diagram of (a) one-phonon absorption process for rotating 
the second spin from $m_2$ to $m_2-1$, and (b) one-phonon emission process 
for rotating the third spin from $m_3$ to $m_3-1$ for a five-spin $S=1$ system. 
The wavy lines denote the phonons emitted or absorbed. 
(a) The energy difference $(E_k-E_m)$ equals $2J+D+H$ that is positive, 
and (b) $(E_k-E_m)$ equals $J+D+H$ that is negative
when $-(2J+D) < H < -(J+D)$.}
\label{fig:examples}
\end{figure}

To understand spin relaxation in the $S \ge 1$ model, we assume
that the spin system is weakly linearly coupled to a surrounding phonon 
bath in $d$ dimensions. Since the phonon bath has much shorter 
relaxation time than the spin system, it is assumed that each spin 
is independently coupled to the bath. Then spin relaxation occurs 
through consecutive emission or absorption of a single phonon with
energy that equals the cost of rotating a single spin from $m_i$ to 
$m_i \pm 1$ in a single transition while keeping the rest of the 
spins fixed. Henceforth we call these rotations first-order 
one-phonon processes. 

A phonon bath is described as a collection of simple harmonic oscillators 
so the phonon Hamiltonian is written as
\begin{eqnarray}
{\cal H}_{\mathrm {ph}} &=& \sum_{\vec{q}}
\hbar \omega_{\vec{q}} ( c_{\vec{q}}^{\dagger} c_{\vec{q}} + \frac{1}{2} ),
\:  \:  \: \: \: \: \: \omega_{\vec{q}} = c \: q,
\label{eq:H_ph}
\end{eqnarray}
where $\vec{q}$ is the phonon wave vector, $\omega_{\vec{q}}$ is the 
angular frequency of a harmonic oscillator, $c_{\vec{q}}^{\dagger}$ and 
$c_{\vec{q}}$ are creation and annihilation operators of a phonon with
wave vector $\vec{q}$, and $c$ is the sound velocity in the lattice. 
The following spin-phonon coupling Hamiltonian ${\cal H}_{\mathrm {sp-ph}}$ 
has the simplest form that takes into account all possible first-order 
one-phonon processes.
\begin{eqnarray}
{\cal H}_{\mathrm {sp-ph}} &=& \lambda \sum_{j=1}^N \sum_{\vec{q}} 
\sqrt{ \frac{\hbar}{2NM \omega_{\vec{q}}} } \: \: q 
(S_j^+ c_{\vec{q}}^{\dagger}  + S_j^+ c_{\vec{q}} + \nonumber \\
& & S_j^- c_{\vec{q}}^{\dagger} + S_j^- c_{\vec{q}} ), 
\label{eq:H_sp-ph} 
\end{eqnarray}
where $\lambda$ is a coupling constant, $N$ is the number of unit cells
associated with the phonon bath, $M$ is the mass of the particle in
the unit cell, and $S_j^{\pm}$ are the raising 
and lowering spin operators for site $j$. Here the polarization of the phonons
is not considered for simplicity. The magneto-elastic coupling 
theory \cite{CALL65} suggests that the spin-phonon coupling must be 
proportional to a linear strain tensor,
$\epsilon_{\alpha^{\prime} \alpha}=\nabla_{\alpha^{\prime}} u_{\alpha}$, where 
$u_{\alpha}$ is the $\alpha$ component of the displacement vector $\vec{u}$ and 
$\alpha^{\prime},\alpha \in \{x,y,z\}$. The displacement vector can
be expressed in terms of $c_{\vec{q}}^{\dagger}$ and $c_{\vec{q}}$,
and a Fourier transform is carried out on $\epsilon_{\alpha^{\prime} \alpha}$. 
This explains the dependence of the prefactor of ${\cal H}_{\mathrm {sp-ph}}$
on the wave vector $\vec{q}$. Equation~(\ref{eq:H_sp-ph}) contains the minimum 
number of terms required to rotate the spin vectors via one-phonon emission 
or absorption processes. Notice that the spin system includes 
a nearest-neighbor exchange interaction, in contrast to works reported by 
other groups \cite{LEUE00,CHUD05}. Due to the exchange interaction,
a transition from $m_l$ to $m_l-1$ does not uniquely determine the
sign of the energy difference $E_k - E_m$ between the two states. 
For example, when $m_l=1$, $E_k - E_m$ is positive for 
Fig.~\ref{fig:examples}(a), while $E_k - E_m$ is negative for 
Fig.~\ref{fig:examples}(b). The same rule is applied to
transition from $m_l$ to $m_l+1$. 

Using Fermi's golden rule, within perturbation theory, we 
calculate the transition rate from state 
$|m \rangle=|m_1 \rangle \bigotimes |m_2 \rangle \bigotimes 
\cdot \cdot \cdot \bigotimes |m_l \rangle \bigotimes 
\cdot \cdot \cdot \bigotimes |m_N \rangle$ 
to state $|k \rangle=|m_1 \rangle \bigotimes |m_2 \rangle \bigotimes 
\cdot \cdot \cdot \bigotimes |m^{\prime}_l \rangle \bigotimes 
\cdot \cdot \cdot \bigotimes |m_N \rangle$, where 
$m^{\prime}_l=m_l \pm 1$. States $|m \rangle$ and $|k \rangle$ differ by 
the rotation of a single spin ($m_l \rightarrow m_l \pm 1$) at site $l$. 
We first consider the transition rate $W_{km}$ from $|m \rangle$ to
$|k \rangle$ for emission of one phonon with energy $\hbar \omega_{\vec{q}}$,
as illustrated in Fig.~\ref{fig:examples}(b).
\begin{widetext}
\begin{eqnarray}
W_{km} &=& \frac{2\pi}{\hbar} \sum_{n_{\vec{q}}} \sum_{\vec{q}}
|\langle n_{\vec{q}} + 1, k| {\cal H}_{\mathrm {sp-ph}}| n_{\vec{q}}, 
m \rangle |^2 \:  \rho_{\mathrm {ph}} 
\: \delta(\hbar \omega_{\vec{q}} - (E_m - E_k)), 
\label{eq:W_int01} \\
&=& \frac{2\pi}{\hbar} \sum_{n_{\vec{q}}} \sum_{\vec{q}} 
\frac{\lambda^2 \: \hbar}{2NM \omega_{\vec{q}}} \: \: q^2 (n_{\vec{q}} + 1) 
 \rho_{\mathrm {ph}} | \langle k | (S_l^+ + S_l^-) | m \rangle |^2 
\delta (\hbar \omega_{\vec{q}} - (E_m - E_k) )
\label{eq:W_int02}
\end{eqnarray}
\end{widetext}
where $\rho_{\rm {ph}}$ is the phonon density of states, $n_{\vec{q}}$ 
is the eigenvalue of the phonon number operator, and
$c^{\dagger}_{\vec{q}} | n_{\vec{q}} \rangle 
= \sqrt{n_{\vec{q}} + 1} | n_{\vec{q}} + 1 \rangle$ is used. Here
$E_k$ and $E_m$ are the energies of states $|k \rangle$ and $|m \rangle$
calculated from the spin Hamiltonian, Eq.~(\ref{eq:H_spin}), and the 
energy difference $\Delta E$ is given by
\begin{eqnarray}
\Delta E (m_l \rightarrow m_l \pm 1) &\equiv& E_k - E_m  \nonumber \\
 &=& \mp J \sum_{k \in{nn(l)}} m_k^{(l)}  \mp D(2m_l \pm 1) \mp H ,
\label{eq:dE_1}
\end{eqnarray}
where the sum runs over nearest neighbors of site $l$.
Using the Bose-Einstein distribution function, one knows that
\begin{eqnarray}
& & \sum_{n_{\vec{q}}} (n_{\vec{q}} + 1)  \rho_{\mathrm {ph}} 
 = \frac{1}{1-e^{-\beta \hbar \omega_{\vec{q}}}}
\label{eq:ph_density}
\end{eqnarray}
where $k_B$ is the Boltzmann constant, $T$ is the temperature,
and $\beta=1/(k_B T)$.
Assuming that the bath relaxes much faster than the spin system, 
we integrate over all degrees of freedom of the bath 
and convert $\sum_{\vec{q}}$ into $[(Na^d)/(2\pi)^d] \int{d^d q}$, 
where $a$ is the lattice spacing. Then the transition rate from state 
$|m \rangle$ to $|k \rangle$ becomes
\begin{eqnarray}
W_{km} &=& \frac{\lambda^2 \: \tilde{N}}{\gamma \eta \hbar^{d+1} 
c^{d+2}} \: \: \frac{(E_m - E_k)^d}{1-e^{-\beta (E_m - E_k)}} , 
\: \: \: \: \: \: E_m  - E_k > 0
\label{eq:W_1st_p} \\
\tilde{N} &=& (S+m_l)(S-m_l+1) \delta_{k_l,m_l-1} \nonumber \\
& & + (S-m_l)(S+m_l+1) \delta_{k_l,m_l+1}, \\
\gamma &=& 2 \pi \: \: (d=3), \:\: \: \: \: 2 \: \: (d=1,2), 
\label{eq:gamma} 
\end{eqnarray}
where $\eta$ is a mass density associated with the bath 
and $k_l$ is the quantum number of
the $l$th spin for state $|k \rangle$. The transition rate for
absorption analogously becomes
\begin{eqnarray}
W_{km} &=& \frac{\lambda^2 \: \tilde{N}}{\gamma \eta \hbar^{d+1} 
c^{d+2}} \: \:  \frac{(E_k - E_m)^d}{e^{\beta (E_k - E_m)}-1} ,
\: \: \:  E_k - E_m > 0. 
\label{eq:W_1st_m}
\end{eqnarray} 
Henceforth, we refer to the derived transition rates, 
Eqs.~(\ref{eq:W_1st_p}) and (\ref{eq:W_1st_m}), as phonon-assisted 
transition rates. 

\section{Consequences of phonon-assisted transition rates} 
The derived phonon-assisted transition rates are generalized forms 
of those for the Ising model discussed in 
Refs.\onlinecite{parknov1,parknov2,parknov3}.
According to Eqs.~(\ref{eq:W_1st_p}) and (\ref{eq:W_1st_m}),
the transition rates are highest for $m_l=0$ and lowest for $m_l=\pm S$ 
for large $S$. A transition rate \cite{CHUD05} similar to the derived 
rates was obtained for an {\it isolated} spin cluster embedded in 
a lattice instead of interacting spin clusters. 
In this formalism, the single-ion anisotropy parameter $D$ corresponds
to the coupling constant $\lambda$ in Eq.~(\ref{eq:H_sp-ph}). 
Assuming that $\lambda \sim D$, 
we estimate the magnitude of the prefactor of the $d=3$ phonon-assisted
transition rates, Eqs.~(\ref{eq:W_1st_p}) and (\ref{eq:W_1st_m}), 
for example, for the nanoscale single-molecule magnet Mn$_{12}$. Using 
measured parameter values such as $S=10$, $m_l=10$, $J=0.01$~K, 
$\eta=1.83 \times 10^3$ kg/m$^2$ [Ref.\onlinecite{LIS80}], and 
$c=1.45 \times 10^3$ m/s, \cite{LEUE00} we find the prefactor to be 
0.00041 s$^{-1}$. Here we use $J=0.01$~K due to the large intermolecular
separation in this system, although it was not directly measured.
To associate the derived transition rates with magnetization relaxation 
times for various nanoscale systems, one needs to solve a master equation 
including the derived transition rates or perform kinetic Monte Carlo 
simulations with the rates. Hereafter we focus on the latter approach.

\begin{figure}
\includegraphics[angle=0,width=0.4\textwidth]{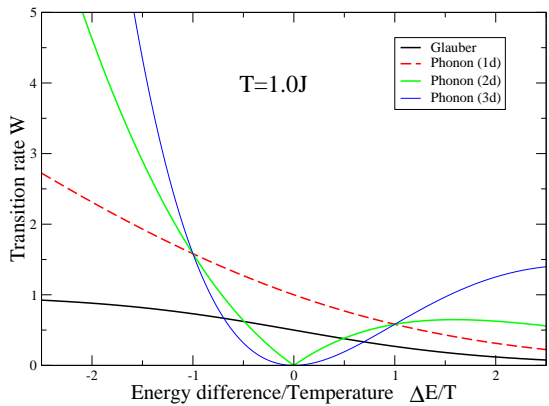}
\caption{Transition rates $W$ vs $\Delta E/T$ for the Glauber and 
phonon-assisted transition rates in the case of $d=1,2$, and 3 
dimensional phonon baths, computed at temperature $T=1.0J/k_B$. 
The prefactor of the phonon-assisted transition rates was 
not included.}
\label{fig:W_T1d0}
\end{figure}

In Fig.~\ref{fig:W_T1d0} the phonon-assisted transition rates for 
$d=1,2$, and 3 dimensional phonon baths are shown as functions of 
$\Delta E/T$ at temperature $T=1.0J$ and compared to the Glauber 
transition rate, $1/(e^{\beta (E_k - E_m)} + 1)$. The main difference 
between the phonon-assisted and Glauber transition rates originates 
from the nature of the bath coupled to the spin system. When degenerate
states are involved in transitions, this difference becomes prominent.
In the $d=2$ and 3 phonon-assisted transition rates, transitions between
degenerate states are forbidden because $W_{km}=0$ when $\Delta E=0$.
In the case of $d=1$, the rate does not vanish for $\Delta E=0$, but 
rather decreases monotonously with $\Delta E/T$ like the Glauber
transition rate. Ramifications of the forbidden transitions on 
dynamic and equilibrium properties are discussed in Monte Carlo 
simulations. For simplicity, we consider a ferromagnetic $S=1$ 
model [the Blume-Capel model with $D>0$ in the spin Hamiltonian,
Eq.~(\ref{eq:H_spin})] on a $L \times L$ square lattice 
with the $d=2$ phonon-assisted transition rates unless specified 
otherwise.

\subsection{Dynamic properties}

\begin{figure}
\includegraphics[angle=0,width=0.4\textwidth]{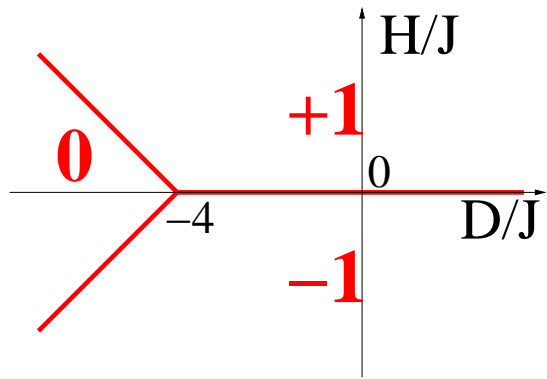}
\caption{Phase diagram of the Blume-Capel model with $D>0$ in
the spin Hamiltonian~(\ref{eq:H_spin}) at zero temperature.
\cite{CIRI96} The equilibrium spin
configurations are shown in the three different regions.}
\label{fig:BC}
\end{figure}

\begin{figure}
\includegraphics[angle=0,width=0.45\textwidth]{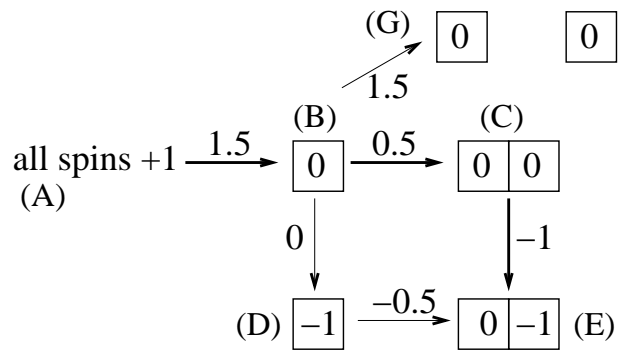}
\caption{Schematic diagram of relaxation of magnetization for
a ferromagnetic $S=1$ Blume-Capel model at $H=-3.25J$ and $D=0.75J$, via
first-order one-phonon processes, $m_l \rightarrow m_l \pm 1$. 
State (A) denotes all spins aligned along the $+z$ axis, and
state (B) a single $m_l=0$ spin in the sea of $m_i=+1$ spins.
The boxes represent rotated spins from the $+z$ axis in the
initial state. State (C) represents two nearest-neighbor 
$m_l=0$ spins in the sea of $m_i=+1$ spins, while state (G) 
represents two $m_l=0$ spins that are not nearest neighbors.
The numbers next to the arrows indicate the energy difference 
$\Delta E$ defined by Eq.~(\ref{eq:dE_1}). With the $d=2$ 
phonon-assisted transition rates, the spin system relaxes 
through (A)$\rightarrow$(B)$\rightarrow$(C)$\rightarrow$(E) 
rather than (A)$\rightarrow$(B)$\rightarrow$(D)$\rightarrow$(E).
The thick arrows represent the most probable path for
relaxation with the phonon-assisted rates. The thin arrows
denote the most probable path with the Glauber transition rate.}
\label{fig:m+1}
\end{figure}

We investigate the effects of the forbidden transitions on the lifetime of
the metastable state for the $S=1$ Blume-Capel model below the critical 
temperature. The equilibrium spin configurations for the model are shown
in different regions in Fig.\ref{fig:BC}.\cite{CIRI96} The nucleation
and metastability for the model were studied for $ -4J < D < -3J$ and
$-J < H < -(4J+D)$ in Ref.\onlinecite{CIRI96}. The critical 
temperature $T_c$ increases as $D/J$ increases at $H$=0. The value of 
$T_c$ at $D=0$ was calculated using different methods. Bethe-lattice 
approximation gave rise to 2.065$J/k_B$ \cite{TANA81} and the effective 
field theory suggested 1.952$J/k_B$.\cite{POLA03} Expanded Bethe-Peierls 
approximation produced 1.915$J/k_B$,\cite{DU03} and Monte Carlo simulations 
suggested that the critical temperature at $H$=0 is 1.6950$J/k_B$ at $D=0$ 
and 2.1855$J/k_B$ at $D=5J$.\cite{SILV02} Suppose that all spins are 
initially aligned along the $+z$ axis. When an external magnetic field 
is applied along the $-z$ axis, the initial state becomes metastable.
When $-(4J+D) < H < 0$, at low temperatures, the spin system relaxes 
toward the stable state (all spins along the $-z$ axis) via creating 
a single critical droplet consisting of connected $m_i=-1$ spins. 
(In the regime studied in Ref.\onlinecite{CIRI96} multiple critical 
droplets are formed.) To give a specific example, we consider 
$H=-3.25J$, $D=0.75J$, and $T < 0.02J/k_B$.
Figure~\ref{fig:m+1} illustrates a few possible relaxation paths from the 
metastable state. The initial state (A) can first relax to state
(B), which represents a single $m_l=0$ spin in the sea of $m_l=+1$ spins. 
Then state (B) can relax to one of the states (C), (D), or (G), 
where state (C) denotes a single $m_l=-1$ spin in the sea of $m_l=+1$ 
spins and state (D) [state (G)] two nearest-neighbor $m_l=0$ spins 
[two $m_l=0$ spins that are not nearest neighbors] in the sea of $m_l=+1$ spins.
Transitions between states (B) and (D) are forbidden because 
$\Delta E= \pm (4J-D+H)=0$. Thus, the system can relax from state (B) 
to either state (C) or (G). The most likely path among the alternative ones 
is (B)$\rightarrow$(C)$\rightarrow$(E), as indicated by the thick arrows
in Fig.~\ref{fig:m+1}. Thus, for the $d=2$ phonon-assisted transition 
rate, the critical droplet is state (C). As a result, the energy barrier 
to be overcome in order to reach the stable state, is 2.0$J$, while 
it is 1.5$J$ for the Glauber transition rate. Accordingly, the relaxation 
time becomes longer than that for the Glauber transition rate.
In an Ising system coupled to a phonon bath, a similar behavior has 
been found at $H=-2J$.\cite{parknov1,parknov2,parknov3} If there are no alternative paths, 
the relaxation time diverges. For a two-electron state in a quantum dot, 
the measured spin-lattice relaxation time was observed to diverge at 
a particular magnetic field where the triplet and the singlet states 
became degenerate.\cite{MEUN07}

\subsection{Equilibrium properties}

\begin{figure}
\includegraphics[angle=0,width=0.3\textwidth]{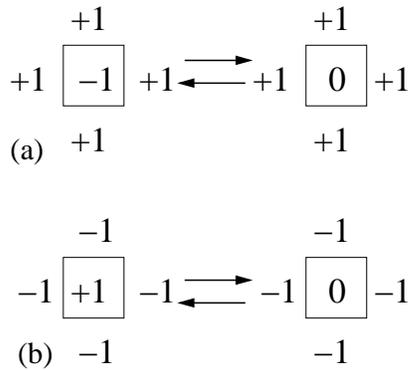}
\caption{Transitions forbidden by first-order one-phonon 
processes for a $S=1$ spin system on a square $L \times L$ 
lattice with $D=4J$ and $H=0$ in which $W_{km}=0$.
(a) Initial state of $-1$ with nearest neighbors of $+1$.
(b) Initial state of $+1$ with nearest neighbors of $-1$.}
\label{fig:forbidden}
\end{figure}

\begin{figure}
\includegraphics[angle=0,width=0.2\textwidth]{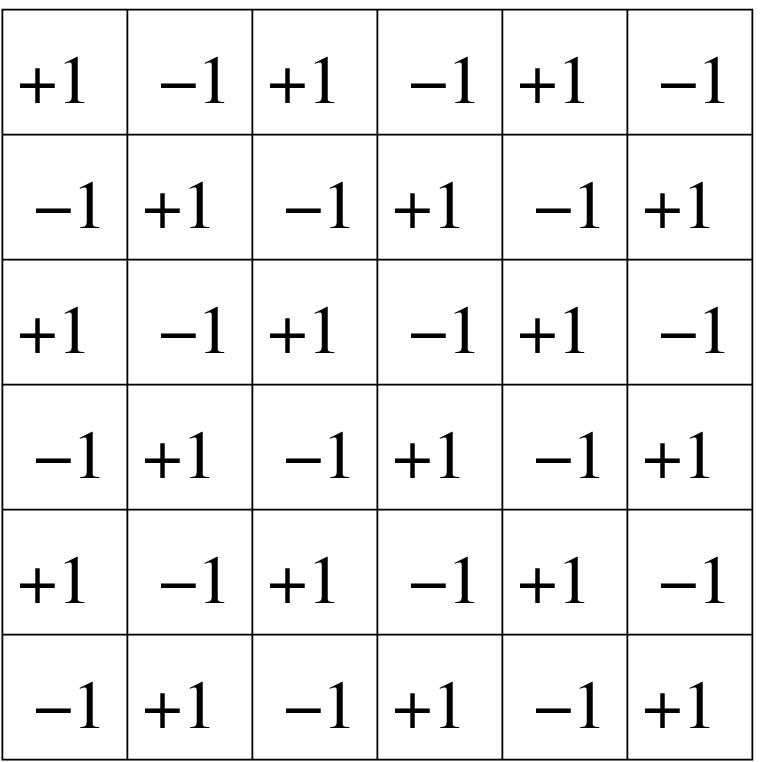}
\caption{A checkerboard state for a $S=1$ spin system
on a $6 \times 6$ square lattice.}
\label{fig:checkerboard}
\end{figure}

The $d=1,2$, and 3 phonon-assisted transition rates satisfy detailed 
balance. Thus, equilibrium properties obtained using the derived 
transition rates must agree with those obtained using different 
transition rates in Monte Carlo simulations. However, there is a caveat 
in this statement because of forbidden transitions between degenerate 
states for the $d=2$ and 3 phonon-assisted rates. In some cases the 
forbidden transitions would prevent the spin system from relaxing 
to the equilibrium state if we start with a particular initial 
state. As an example, we consider $D=4J$ and $H=0$. As illustrated 
in Fig.~\ref{fig:forbidden}, transitions between $m_l=+1$ ($m_l=-1$) 
and $m_l=0$ with the sum of the nearest neighbors fixed as $-4$ ($+4$) 
are not allowed because $\Delta E = -4 J \pm H + D=0$. 
So if we started with a checkerboard initial state as shown in 
Fig.~\ref{fig:checkerboard} in Monte Carlo simulations, 
the system would stay indefinitely at the initial state 
because $W_{km}=0$ for any possible single-spin rotations. 
However, we have confirmed that the system reaches equilibrium 
if we start with a random initial state or a state slightly modified 
from the checkerboard pattern. (Even one defect site in the perfect 
checkerboard state is sufficient.) A similar feature was reported 
in the time evolution of field-driven solid-on-solid interfaces 
using the $d=2$ phonon-assisted transition rate.\cite{BUEN07}

\begin{figure}
\includegraphics[angle=0,width=0.45\textwidth]{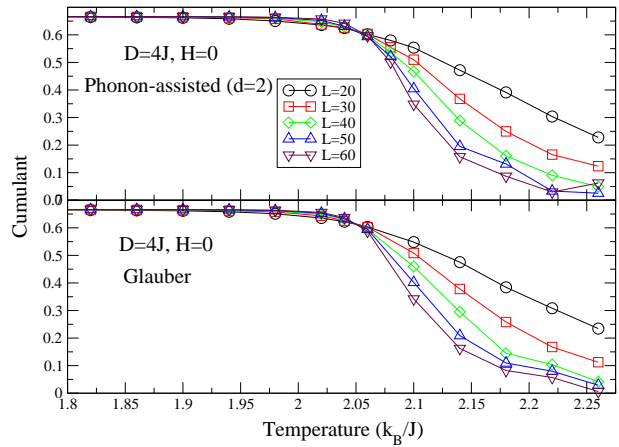}
\caption{Cumulant $U_L$ [Eq.~(\ref{eq:UL})] vs temperature calculated 
from Monte Carlo simulations using (top) the Glauber and (bottom) 
the $d=2$ phonon-assisted transition rates for $S=1$ spins 
on a $L \times L$ square lattice at 
$D=4J$ and $H=0$. Random initial states and periodic boundary 
conditions were used. For the fixed point of the cumulant and
critical temperature, see the text.}
\label{fig:UL_D4}
\end{figure}

Monte Carlo simulations were performed for $D=4J$ and $H=0$ with 
periodic boundary conditions and a random initial state.
$L=20, 30, 40, 50$, and 60 were considered at several different
temperatures. The average absolute magnetization per site 
$\langle |m| \rangle$ and cumulant $U_L$ were calculated. 
\begin{eqnarray}
\langle |m| \rangle &=& \frac{1}{\tilde{M}} \sum_{i=1}^{\tilde{M}} 
\frac{1}{L^2} \left| \sum_{j=1}^{L^2} S_j^{(i)} \right|, \: \: \: 
U_L = 1 - \frac{\langle m^4 \rangle}{3 \langle m^2 \rangle^2},
\label{eq:UL}
\end{eqnarray}
where $m$ is the magnetization per site and ${\tilde M}$ is the total 
number of Monte Carlo steps. It was found that $\langle |m| \rangle$
(not shown) and $U_L$ (Fig.~\ref{fig:UL_D4}) as functions of
temperature agree with those obtained using the Glauber transition
rate for the different system sizes. For the Glauber transition rate,
the values of $U_L$ for the different system sizes intersect with 
one another at $U_L^{\star}$=0.611$\pm$ 0.008 and $T_c$=2.052$\pm$0.005
$J/k_B$, while for the phonon-assisted transition rate we have 
$U_L^{\star}$=0.611$\pm$ 0.009 and $T_c$=2.052$\pm$0.006 $J/k_B$.
So both transition rates give rise to the same critical temperature.
This critical temperature is slightly lower than that reported in 
Ref.\onlinecite{SILV02}, 2.13$J/k_B$.

\section{Higher-order processes}

We have, so far, discussed spin relaxation caused by first-order 
one-phonon processes, $m_l \rightarrow m_l \pm 1$. However, 
higher-order processes such as multi-phonon processes and 
second-order one-phonon processes ($m_l \rightarrow m_l \pm 2$), 
can also contribute to the spin relaxation. Their contributions
become significant especially when encountering transitions 
forbidden by the first-order one-phonon processes. In this case,
the spin system would find less costly paths via the higher-order 
processes than paths directed by the first-order processes.
Thus, the relaxation time becomes shortened and the system may 
recover the same relaxation time or energy barrier as the Glauber 
transition rate. The system starting with the checkerboard state 
(Fig.~\ref{fig:checkerboard}) can be also relaxed to equilibrium.

Recently, multi-phonon processes such as Raman processes and 
two-phonon processes were considered in a spin-phonon relaxation 
rate for rigid atomic clusters, and it was shown that there are 
no closed analytical forms for the rate.\cite{CALE06}
In the current study, we focus on second-order one-phonon 
processes as higher-order processes. These second-order processes
were included as a part of the relaxation mechanism
for the single-molecule magnet Mn$_{12}$.\cite{LEUE00} The simplest 
form of the spin-phonon coupling Hamiltonian for these processes 
is given by
\begin{eqnarray}
{\cal H}_{\mathrm {sp-ph}}^{\rm{(2nd)}} &=& \lambda^{\prime} 
\sum_{j=1}^N \sum_{\vec{q}} \sqrt{ \frac{\hbar}{2NM \omega_{\vec{q}}} } 
\: q [ (S_j^+)^2 c_{\vec{q}}^{\dagger} \nonumber \\
& & + (S_j^-)^2 c_{\vec{q}}^{\dagger} 
+ (S_j^+)^2 c_{\vec{q}} + (S_j^-)^2 c_{\vec{q}} ],
\label{eq:H_sp-ph-2nd} 
\end{eqnarray}
where $\lambda^{\prime}$ is a coupling constant and 
$|\lambda^{\prime}| \ll |\lambda|$. Following the method used 
in Sec.~II, we obtain the transition rate $W_{km}^{\mathrm{(2nd)}}$ 
from state $|m \rangle$ to $|k \rangle$, where these two states
differ by a single spin rotation at site $l$, $m^{\prime}_l=m_l \pm 2$.
\begin{equation}
W_{km}^{\mathrm{(2nd)}}=\frac{(\lambda^{\prime})^2 \: N^{\prime}}
{\gamma \eta \hbar^{d+1} c^{d+2}}
\: \: \left| \frac{(E_k - E_m)^d}{e^{\beta (E_k - E_m)}-1} \right|, 
\label{eq:W_2nd} 
\end{equation}
\begin{eqnarray}
& & N^{\prime}=(S+m_l)(S-m_l+1)(S+m_l-1)(S-m_l+2) \delta_{k_l,m_l-2} 
\nonumber \\
 & & + (S-m_l)(S+m_l+1)(S-m_l-1)(S+m_l+2) \delta_{k_l,m_l+2},  
\end{eqnarray}
where $\gamma$ is defined in Eq.~(\ref{eq:gamma}).
The energy difference $E_k-E_m=\Delta E$ is 
\begin{eqnarray}
\Delta E (m_l \rightarrow m_l \pm 2) &=&
 \mp 2 J \sum_{k \in{nn(l)}} m_k^{(l)}  \mp 4 D(m_l \pm 1) \nonumber \\
& & \mp 2 H,
\label{eq:dE_2}
\end{eqnarray}
where the sum runs over nearest neighbors of site $l$.
This formula is applied to both emission and absorption
processes.

\begin{figure}
\includegraphics[angle=0,width=0.45\textwidth]{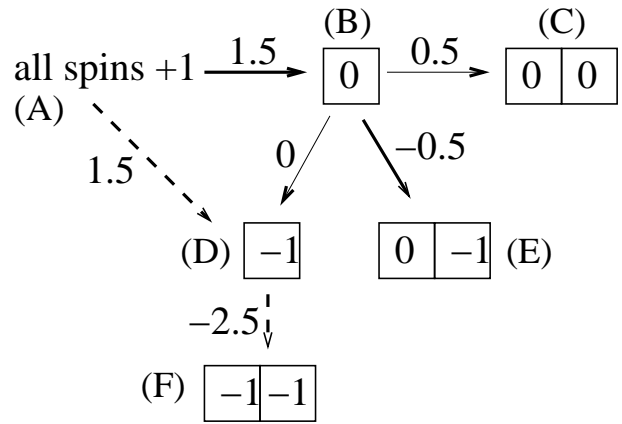}
\caption{Schematic diagram of relaxation of magnetization for
a $S=1$ spin system at $H=-3.25J$ and $D=0.75J$ when both 
first-order ($m_l \rightarrow m_l \pm 1$) and second-order 
one-phonon processes ($m_l \rightarrow m_l \pm 2$) are
considered. The thick solid and dashed arrows represent
two highly probable paths for relaxation. 
Compared to Fig.~\ref{fig:m+1}, the spin system can relax 
from state (A) to state (E) through (B),
or to state (F) through (D). The numbers next to the arrows indicate 
the energy difference $\Delta E$ defined by Eqs.~(\ref{eq:dE_1}) and 
(\ref{eq:dE_2}).}
\label{fig:m+2}
\end{figure}

When the second-order transition rate $W_{km}^{\mathrm{(2nd)}}$ is 
included in the calculation of the lifetime of the metastable
state, relaxation scenarios are greatly modified as 
illustrated in Fig.~\ref{fig:m+2}. 
With the same parameter values used in Sec.III.A ($D=0.75J$
and $H=-3.25J$), the system can now relax through transitions
as indicated by the thick dashed arrows 
[(A)$\rightarrow$(D)$\rightarrow$(F)] or by the thick solid 
arrows [(A)$\rightarrow$(B)$\rightarrow$(E)] in Fig.~\ref{fig:m+2}. 
None of these transitions involve degenerate states. The critical
droplet for the first relaxation route is state (D), while that
for the second route is state (B). In both relaxation paths, 
the energy barrier is $1.5J$, which is the same as that for the 
Glauber transition rate. In the case of equilibrium Monte Carlo
simulations, the second-order processes allow the system to
relax via alternative second-order transitions with lower energy
cost but $\Delta E \neq 0$. So the checkerboard initial state
(Fig.~\ref{fig:checkerboard}) can reach equilibrium 
for $D=4J$ and $H=0$. It is confirmed that for $L=20$ the 
equilibrium properties computed with addition of 
$W_{km}^{\mathrm{(2nd)}}$ to $W_{km}$ agree with 
those obtained using the Glauber transition rate.

\section{Conclusion}

We have considered the $S \ge 1$ model weakly coupled to 
a one-, two-, or three-dimensional phonon bath and derived corresponding 
transition rates from the spin-phonon coupling Hamiltonian. 
The derived phonon-assisted transition rates for two- and three- 
dimensional baths differ from other transition rates in that
the former rates become zero for degenerate states. This caused
some transitions to be forbidden by the first-order one-phonon 
assisted transition rates, increasing the magnetization relaxation 
time. Using a combination of the first-order one-phonon processes 
with the second-order processes, however, the system found more 
energy-efficient paths to equilibrium, and the relaxation time 
shortened. These results represent a major step toward developing
physically realistic kinetic Monte Carlo simulations for magnetic
spin systems.

\section*{Acknowledgments}
The author is grateful to M. A. Novotny, L. Solomon,
and P. A. Rikvold for discussions.


\end{document}